\documentclass[aps,pre,twocolumn,showpacs,superscriptaddress]{revtex4}
\usepackage{graphicx,color,epsfig}
\begin{document}
\title{Local versus global interactions in nonequilibrium transitions:
A model of social dynamics}
\author{J. C. Gonz\'alez-Avella}
\affiliation{IMEDEA (CSIC-UIB), Campus Universitat Illes Balears,
E-07122 Palma de Mallorca, Spain}
\author{V. M. Egu\'iluz}
\affiliation{IMEDEA (CSIC-UIB), Campus Universitat Illes Balears,
E-07122 Palma de Mallorca, Spain}
\author{M. G. Cosenza}
\affiliation{IMEDEA (CSIC-UIB), Campus Universitat Illes Balears,
E-07122 Palma de Mallorca, Spain} \affiliation{Centro de F\'isica
Fundamental, Universidad de los Andes, Apartado Postal 26, M\'erida
5251, Venezuela}
\author{K. Klemm}
\affiliation{Bioinformatics, Dept.\ of Computer Science,
University of Leipzig, H\"artelstr.\ 16, 04107 Leipzig, Germany}
\author{J. L. Herrera}
\affiliation{Centro de F\'isica Fundamental, Universidad de los
Andes, Apartado Postal 26, Merida 5251, Venezuela}
\author{M. San Miguel}
\affiliation{IMEDEA (CSIC-UIB), Campus Universitat Illes Balears,
E-07122 Palma de Mallorca, Spain}
\date{\today}
\begin{abstract}
A nonequilibrium system of locally interacting elements in a
lattice with an absorbing order-disorder phase transition is
studied under the effect of additional interacting fields. These
fields are shown to produce interesting effects in the
collective behavior of this system. Both for autonomous and
external fields, disorder grows in the system when the probability
of the elements to interact with the field is increased. There
exists a threshold value of this probability beyond which the
system is always disordered. The domain of parameters of the
ordered regime is larger for nonuniform local fields than for
spatially uniform fields. However, the zero field limit is discontinous.
In the limit of vanishingly
small probability of interaction with the field, autonomous or
external fields are able to order a system that would fall in a
disordered phase under local interactions of the elements alone.
We consider different types of fields which are interpreted as
forms of mass media acting on a social system in the context of
Axelrod's model for cultural dissemination.
\end{abstract}
\maketitle

\section{Introduction}

The emergence of nontrivial collective behavior in spatiotemporal
dynamical systems is a central issue in the current research on
complex systems, as in many physical, chemical, biological,
economic and social phenomena. There are a variety of processes
occurring in these systems where both spatially local and global
interactions extending all over the system coexist and contribute
in different and competing ways to the collective dynamics. Some
examples include Turing patterns \cite{Turing} (with slow and fast
diffusion), Ginzburg-Landau dynamics \cite{Mikhailov}, surface
chemical reactions \cite{Mertens}, sand dunes (with the motions of
wind and of sand) \cite{Ouchi}, and pattern formation in some
biological systems \cite{Meinhart}. Recently, the collective
behavior of dynamical elements subject to both local and global
interactions has been experimentally investigated in arrays of
chaotic electrochemical cells \cite{Hudson3}. Many of these
systems can be modeled as networks of coupled dynamical units
with coexisting local and global interactions \cite{Kaneko}.
Similarly, the phenomena of pattern formation and collective
behavior induced by external forcing on spatiotemporal systems,
such as chemical reactions \cite{Swinney,Hudson1} or granular
media \cite{Swinney1}, has also been considered. The analogy
between external forcing and global coupling in spatiotemporal
dynamical systems has recently been explored in the framework of
coupled map lattice models \cite{CPP,PC}. It has been found that,
under some circumstances, the collective behavior of an autonomous
spatiotemporal system with local and global interactions is
equivalent to that of a driven spatiotemporal system possessing
similar local couplings as in the autonomous system.

The addition of a global interaction to a locally coupled system
is known to be able to induce phenomena not present in that
system, such as chaotic synchronization and new spatial patterns.
However, the classification and description of generic effects
produced by external fields or global coupling in a nonequilibrium
system of locally interacting units is still an open general
question. The common wisdom for equilibrium systems is that under
a strong external field, local interactions become negligible, and
the system orders following the external field. For nonequilibrium
nonpotential dynamics
\cite{nonpotential} this is not
necessarily the case, and nontrivial effects might arise depending
on the dynamical rules.

This problem is, in particular, relevant for recent studies of
social phenomena in the general framework of complex systems. The
aim is to understand how collective behaviors arise in social
systems. Several mathematical models, many of them based on
discrete-time and discrete-space dynamical systems, have been
proposed to describe a variety of phenomena occurring in social
dynamics
\cite{Weidlich,Ball,Stauffer,minority,Redner,EPJB,Galam,PD,MaxiR}.
In this context, specially interesting is the lattice model
introduced by Axelrod \cite{Axelrod}  to investigate the
dissemination of culture among interacting agents in a society
\cite{Castellano,Shibanai,MaxiR,Vilone,Maxi1,Maxi2,Maxi3,Maxi4}. The state
of an agent in this model is described by a set of individual
cultural features. The local interaction between neighboring
agents depends on the cultural similarities that they share and
similarity is enhanced as a result of the interaction. From the
point of view of statistical physics, this model is appealing
because it exhibits a nontrivial out of equilibrium transition
between an ordered phase (a homogeneous culture) and a disordered
(multicultural) one,  as in other well studied lattice systems
with phase ordering properties \cite{Marro}. The additional effect
of global coupling in this system has been considered as a model
of influence of mass media \cite{Shibanai}. It has also been shown
that the addition of external influences, such as random
perturbations \cite{Maxi2} or a fixed field \cite{GCT}, can induce
new order-disorder nonequilibrium transitions in the collective
behavior of Axelrod's model. However, a global picture of the
results of the competition between the local interaction among the
agents and the interaction through a global coupling field or an
external field is missing. In this paper we address this general
question in the specific context of Axelrod´s model.

We deal with states of the elements of the system and interacting
fields described by vectors whose components can take discrete
values. The interaction dynamics of the elements among themselves
and with the fields is based on the similarity between state
vectors, defined as the fraction of components that these vectors
have in common. We consider interaction fields that originate
either externally (an external forcing) or from the contribution
of a set of elements in the system (an autonomous dynamics) such
as global or partial coupling functions. Our study allows to
compare the effects that driving fields or autonomous fields of
interaction have on the collective properties of systems with this
type of nonequilibrium dynamics. In the context of social
phenomena, our scheme can be considered as a model for a social
system interacting with global or local mass media that represent
endogenous cultural influences or information feedback, as well as
a model for a social system subject to an external cultural
influence. Our results indicate that the usual equilibrium notion
that the application of a field should enhance order in a system
does not hold here. On the contrary, disorder builds-up by
increasing the probability of interaction of the elements with the
field. This occurs independently of the nature (either external or
autonomous) of the field of interaction added to the system.
Moreover, we find that a spatially nonuniform field of interaction
may actually produce less disorder in the system than a uniform
field.

The model, including the description of three types of interaction
fields being considered, is presented in Sec.~II.  In Sec.~III,
the effects of the fields in the ordered phase of the system are
shown, while Sec.~IV analyzes these effects in the disordered
phase. Section~V contains a global picture and interpretation of
our results.

\section{The model}
The system consists of $N$ elements as the sites of a square
lattice. The state $c_i$ of element $i$ is defined as a vector of
$F$ components $\sigma_i=
(\sigma_{i1},\sigma_{i2},\ldots,\sigma_{iF})$. In Axelrod's model,
the $F$ components of $c_i$ correspond to the cultural features
describing the $F$-dimensional culture of element $i$. Each
component $\sigma_{if}$ can take any of the $q$  values in the set
$\{ 0, 1, \ldots, q-1 \}$  (called cultural traits in Axelrod's
model). As an initial condition, each element is randomly and
independently assigned one of the $q^F$ state vectors with uniform
probability. We introduce a vector field $M$ with components
$(\mu_{i1},\mu_{i2},\ldots,\mu_{iF})$. Formally, we treat the
field at each element $i$ as an additional neighbor of $i$ with
whom an interaction is possible. The field is represented as an
additional element $\phi(i)$ such that $\sigma_{\phi(i) f} =
\mu_{if}$ in the definition given below of the dynamics. The
strength of the field is given by a constant parameter $B \in
[0,1]$ that measures the probability of interaction with the
field. The system evolves by iterating the following steps:

(1) Select at random an element $i$ on the lattice (called active
element).

(2) Select the source of interaction $j$. With probability $B$ set
$j=\phi(i)$ as an interaction with the field. Otherwise, choose
element $j$ at random among the four nearest neighbors (the von
Neumann neighborhood) of $i$ on the lattice.

(3) Calculate the overlap (number of shared components)
$l(i,j)=\sum_{f=1}^{F}\delta_{\sigma_{if},\sigma_{jf}}$. If $0<l(i,j)<F$,
sites $i$ and $j$ interact with probability $l(i,j)/F$. In case of interaction,
choose $h$ randomly such that $\sigma_{ih}\neq \sigma_{jh}$ and set
$\sigma_{ih} = \sigma_{jh}$.

(4) Update the field $M$ if required (see definitions of fields
below). Resume at (1).

Step (3) specifies the basic rule of a nonequilibrium dynamics
which is at the basis of most of our results. It has two
ingredients: i) A similarity rule for the probability of
interaction, and ii) a mechanism of convergence to an homogeneous
state.

Before considering the effects of the field $M$, let us review
the original model without field ($B=0$). In any finite network
the dynamics settles into an absorbing state, characterized by
either $l(i,j)=0$ or $l(i,j)=F$, for all pairs of neighbors
$(i,j)$. Homogeneous (''monocultural") states correspond to
$l(i,j)=F$,  $\forall i,j$, and obviously there are $q^F$ possible
configurations of this state.  Inhomogeneous (''multicultural")
states consist of two or more homogeneous domains interconnected
by elements with zero overlap and therefore with frozen dynamics.
A domain is a set of contiguous sites with identical state
vectors. It has been shown that the system reaches ordered,
homogeneous states for $q < q_c$ and disordered, inhomogeneous
states for $q > q_c$, where $q_c$ is a critical value that depends
on $F$ \cite{Castellano,Vilone,Maxi1,Maxi2,Maxi3}. This
order-disorder nonequilibrium transition is of second order in
one-dimensional systems and of first order in two-dimensional
systems \cite{Maxi4}. It has also been shown that the
inhomogeneous configurations are not stable: single feature
perturbations acting on these configurations unfreeze the
dynamics. Under repeated action of these perturbations the system
reaches an homogeneous state \cite{Maxi2}.

To characterize the transition from an homogeneous state to a
disordered state, we consider as an order parameter the average
fraction of cultural domains $g=\langle N_g \rangle /N$. Here
$N_g$ is the number of domains formed in the final state of the
system for a given realization of initial conditions. Figure~1
shows the quantity $g$ as a function of the number of options per
component $q$, for $F=5$, when no field acts on the system
($B=0$). For values of $q<q_c \approx 25$, the system always
reaches a homogeneous state characterized by values $g \rightarrow
0$. On the other hand, for values of $q > q_c$,
the system settles into a disordered state, for which $\langle N_g
\rangle \gg 1$. Another previously used order parameter
\cite{Castellano,Maxi1}, the average size of the largest domain
size, $\langle S_{\rm max}\rangle/N$, is also shown in Fig.~1
for comparison. In this case, the ordered phase corresponds to
$\langle S_{\rm max}\rangle/N =1$, while complete disorder is
given by $\langle S_{\rm max}\rangle/N \rightarrow 0$. Unless
otherwise stated, our numerical results throughout the paper are
based on averages over 50 realizations for systems of size
$N=40\times 40$, and $F=5$.

\begin{figure}[t]
\includegraphics[width=0.45\textwidth,angle=-0]{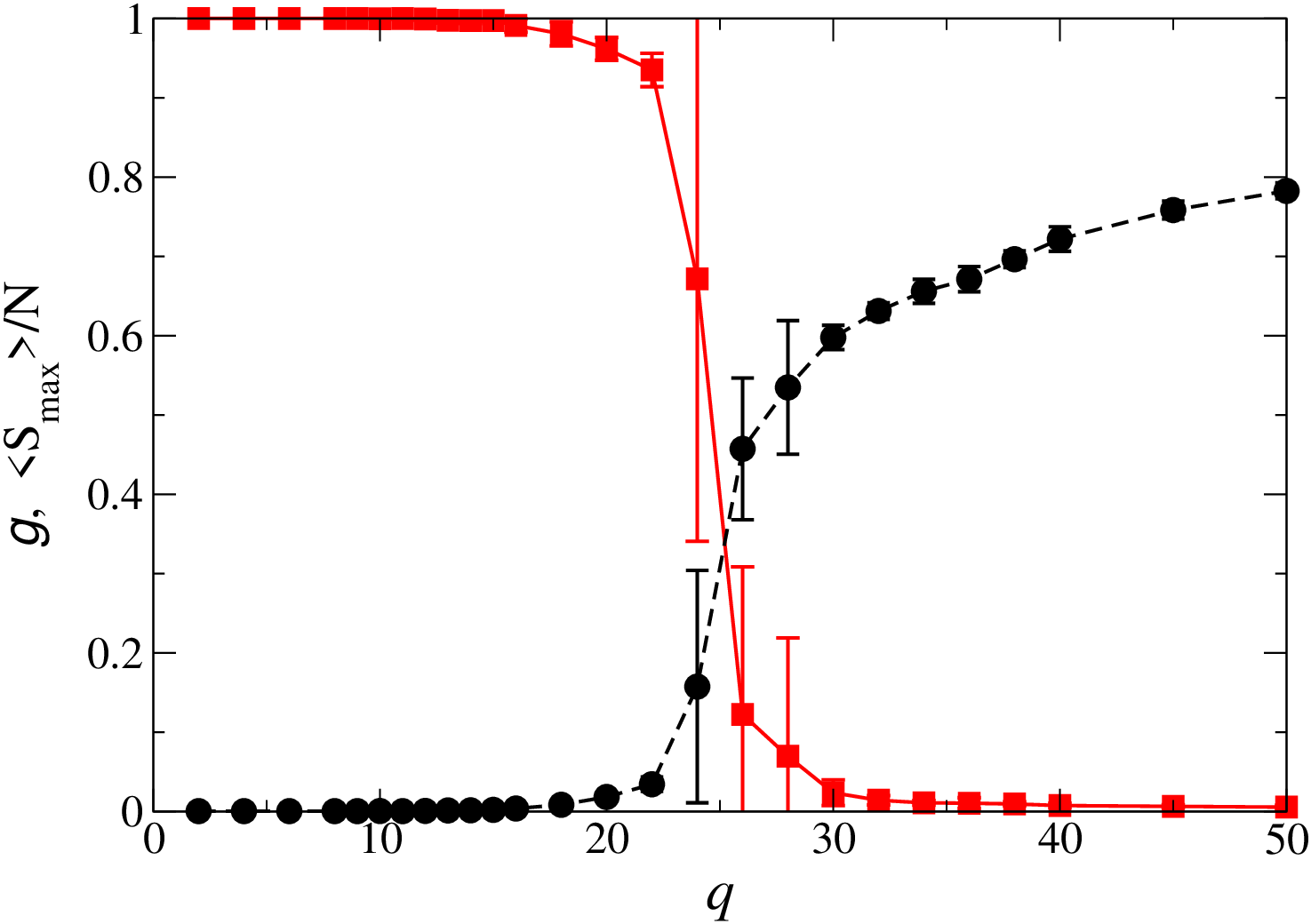}
\caption{Order parameters $g$ (circles) and $\langle
S_{max}\rangle/N$ (squares) as a function of $q$,
in the absence of a field $B=0$.}
\end{figure}

Let us now consider the case where the elements on the lattice have a
non-zero probability to interact with the field ($B>0$). We distinguish
three types of fields.

(i) The \emph{external field} is spatially uniform and constant in
time. Initially for each component $f$, a value $\epsilon_f \in
\{1,\dots,q\}$ is drawn at random and $\mu_{if}=\epsilon_f$ is set
for all elements $i$ and all components $f$. It corresponds to a
constant, external driving field acting uniformly on the system. A
constant \emph{external field} can be interpreted as a specific
cultural state (such as advertising or propaganda) being imposed
by controlled mass media on all the elements of a social system
\cite{GCT}.

(ii) The \emph{global field} is spatially uniform and may vary in
time. Here $\mu_{if}$ is assigned the most abundant value
exhibited by the $f$-th component of all the state vectors in the
system. If the maximally abundant value is not unique, one of the
possibilities is chosen at random with equal probability. This
type of field is a global coupling function of all the elements in
the system. It provides the same global information feedback to
each element at any given time but its components may change  as
the system evolves. In the context of cultural models
\cite{Shibanai}, this field may represent a global mass media
influence  shared identically by all the agents and which contains
the most predominant  trait in each cultural feature present in a
society (a ``global cultural trend").

(iii) The \emph{local field}, is spatially non-uniform and
non-constant. Each component $\mu_{if}$ is assigned the most
frequent value present in component $f$ of the state vectors of
the elements belonging to the von Neumann neighborhood of element
$i$. If there are two or more maximally abundant values of
component $f$ one of these is chosen at random with equal
probability. The \emph{local field} can be interpreted as local
mass media conveying the ``local cultural trend" of its
neighborhood to each element in a social system.

Case (i) corresponds to a driven spatiotemporal dynamical system.
On the other hand, cases (ii) and (iii) can be regarded as
autonomous spatiotemporal dynamical systems. In particular, a
system subject to a \emph{global field} corresponds to a network
of dynamical elements possessing both local and global
interactions. Both the constant \emph{external field} and the
\emph{global field} are uniform. The \emph{local field} is
spatially non-uniform; it depends on the site $i$. In the context
of cultural models,  systems subject to either \emph{local} or
\emph{global fields} describe social systems with endogenous
cultural influences, while the case of the \emph{external field}
represents and external cultural influence.

The strength of the coupling to the interaction field is controlled
by the parameter $B$.
We shall assume that $B$ is uniform, i.e., the field reaches all
the elements with the same probability. In the cultural dynamics
analogy, the parameter $B$ can be interpreted as the probability
that the mass media vector has to attract the attention of the
agents in the social system. The parameter $B$ represents
enhancing factors of the mass media influence that can be varied,
such as its amplitude, frequency, attractiveness, etc.

\begin{figure}[t]
\includegraphics[width=0.45\textwidth]{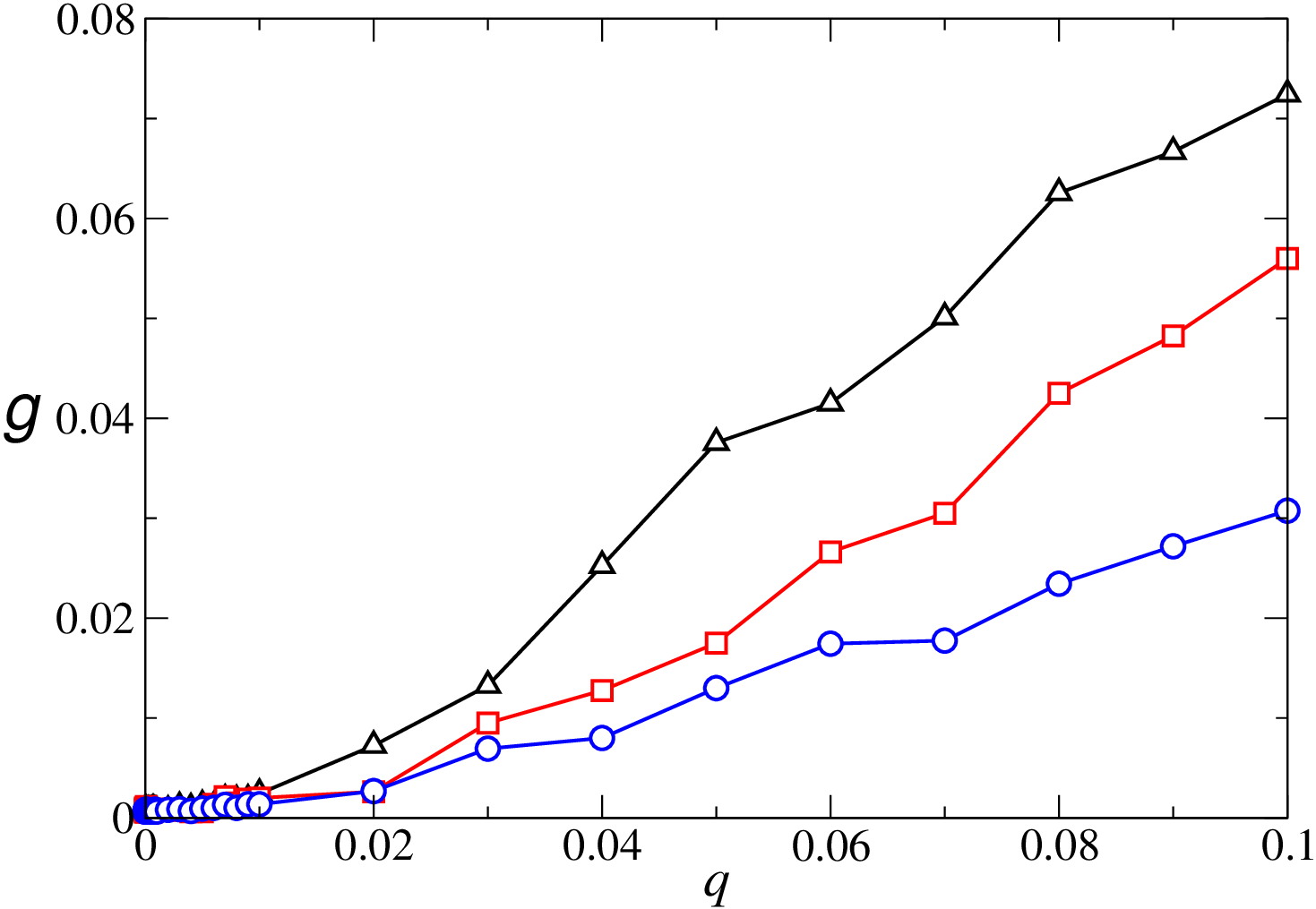}
\caption{Order parameter $g$ as a function of the coupling strength $B$ of
an \emph{external} (squares), \emph{global} (circles)
and \emph{local} (triangles) \emph{field}. Parameter value
$q=10 < q_c$.}
\end{figure}

\section{Effects of an interacting Field for $q<q_c$}

In the absence of any interaction field, the system settles into
one of the possible $q^F$ homogeneous states for $q<q_c$ (see
Fig.~1). Figure~2 shows the order parameter $g$ as a function of
the coupling strength $B$ for the three types of fields. When the
probability $B$ is small enough, the system still reaches in its
evolution a homogeneous state ($g\rightarrow 0$) under the action
of any of these fields. In the case of an \emph{external field},
the homogeneous state reached by the system is equal to the field
vector \cite{GCT}. Thus, for small values of $B$, a constant
\emph{external field} imposes its state over all the elements in
the system, as one may expect. With a \emph{global} or with a
\emph{local field}, however, for small $B$ the system can reach
any of the possible $q^F$ homogeneous states, depending on the
initial conditions. Regardless of the type of field, there is a
transition at a threshold value of the probability $B_c$ from a
homogeneous state to a disordered state characterized by an
increasing number of domains as $B$ is increased. Thus, we find
the counterintuitive result that, above some threshold value of
the probability of interaction, a field induces disorder in a
situation in which the system would order (homogeneous state) under
the effect alone of local interactions among the elements.

\begin{figure}[t]
\includegraphics[width=0.45\textwidth]{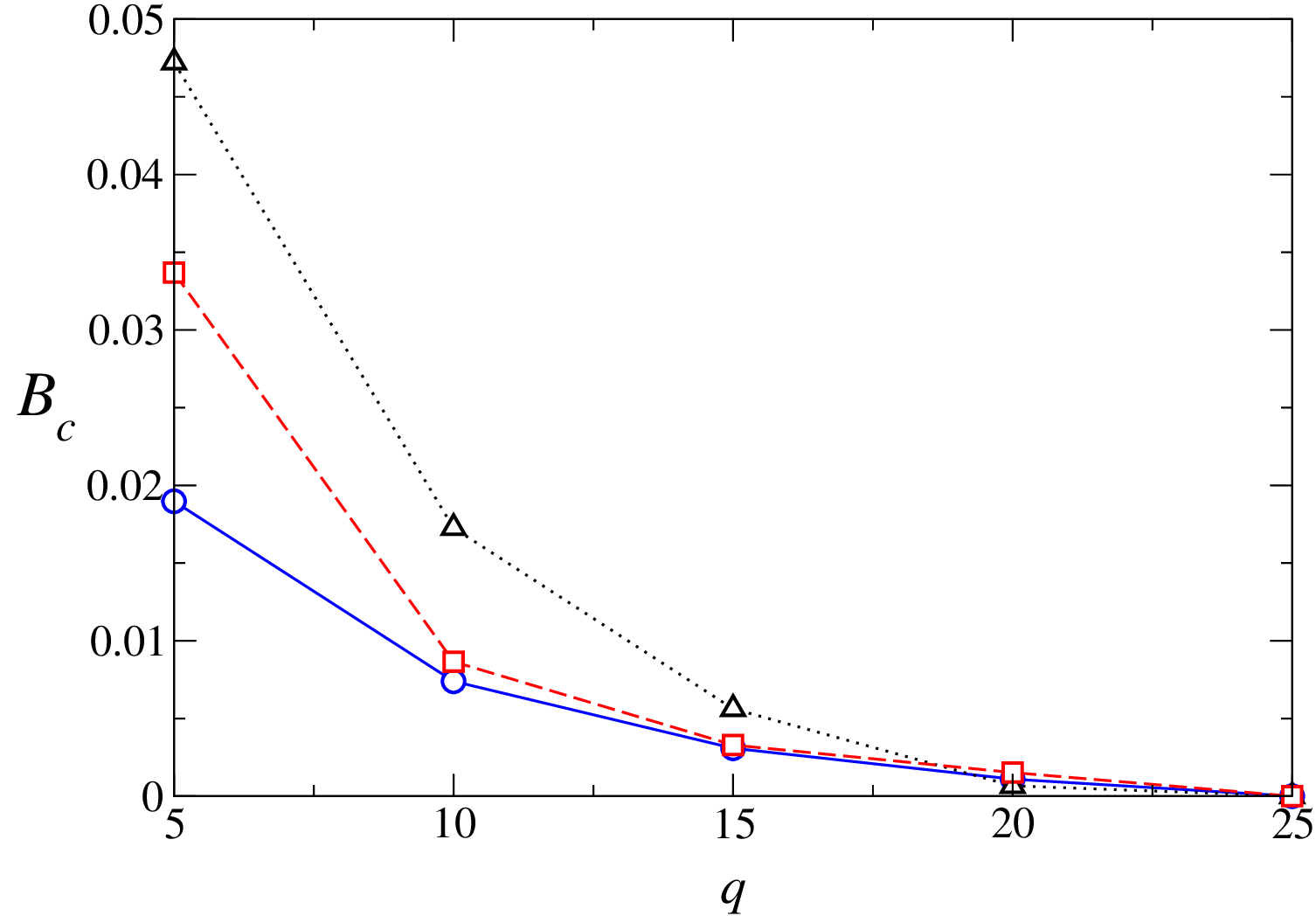}
\caption{Threshold values $B_c$ for $q<q_c$  corresponding to the
different fields. Each line separates the region of order (above
the line)  from the region of disorder (below the line) for an
\emph{external} (squares), \emph{global} (circles), 
and \emph{local} (triangles) \emph{field}.}
\end{figure}

The threshold values of the probability $B_c$ for each type of
field, obtained by a regression fitting \cite{GCT}, are plotted as
a function of $q$ in the phase diagram of Fig.~3. The threshold
value $B_c$ for each field decreases with increasing $q$ for
$q<q_c$. The value $B_c=0$ for the three fields is reached at
$q=q_c\approx 25$, corresponding to the critical value in absence
of interaction fields observed in Fig.~1. For each case, the
threshold curve $B_c$ versus $q$ in Fig.~3 separates the region of
disorder from the region where homogeneous states occur on the
space of parameters $(B,q)$. For $B>B_c$, the interaction with the
field dominates over the local interactions among the individual
elements in the system. Consequently, elements whose states
exhibit a greater overlap with the state of the field have more
probability to converge to that state. This process contributes to
the differentiation of states between neighboring elements and to
the formation of multiple domains in the system for large enough
values of the probability $B$.

\begin{figure}[t]
\includegraphics[width=0.45\textwidth,angle=-0]{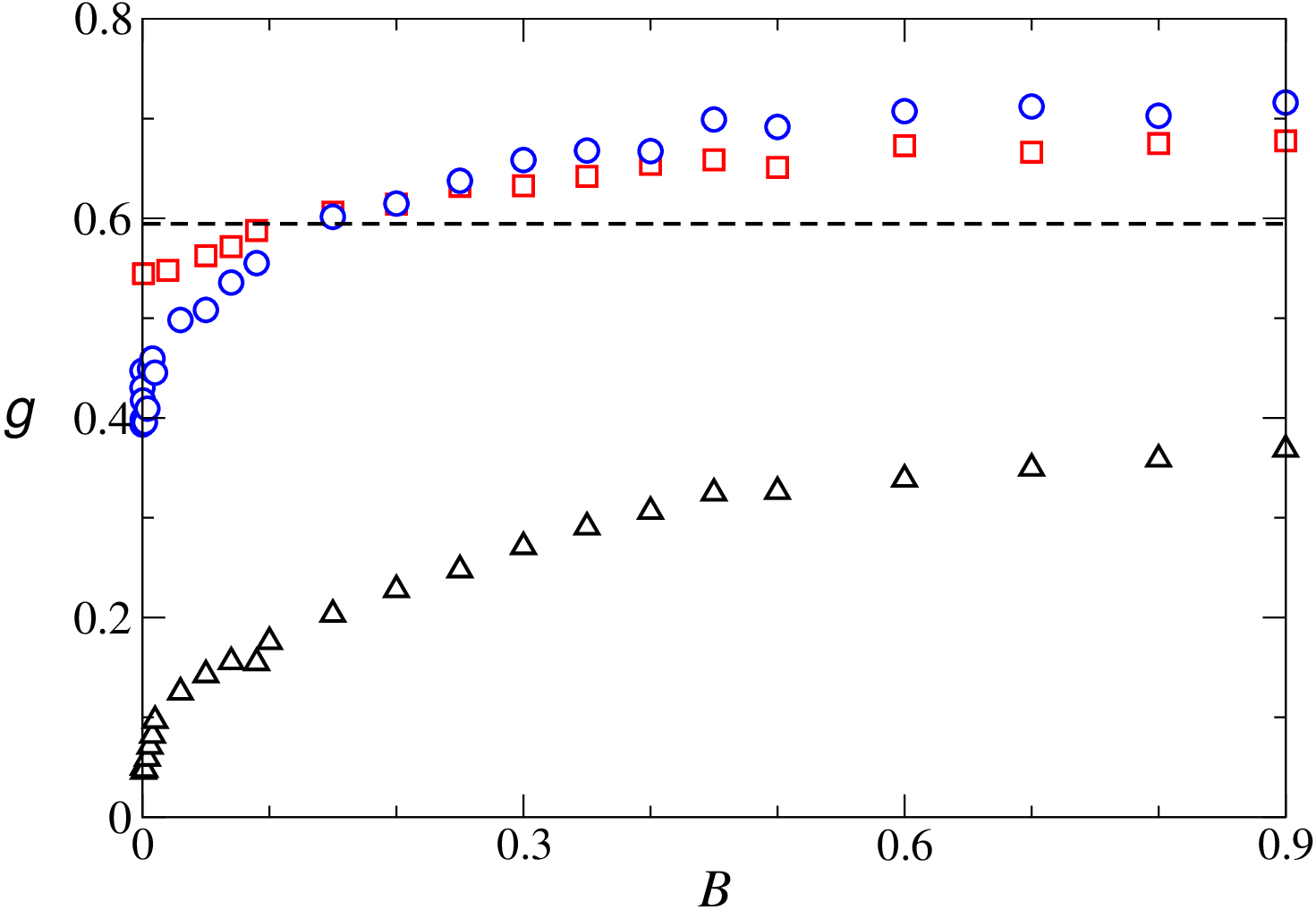}
\caption{Order parameter $g$ as a function of the coupling strength $B$ of
an \emph{external} (squares), \emph{global}
(circles) and \emph{local} (triangles) \emph{field}. The horizontal dashed
line indicates the value of $g$ at $B=0$. Parameter value $q=30$.}
\end{figure}

Note that the region of homogeneous ordered states in the $(B,q)$
space in Fig.~3 is larger for the \emph{local field} than for the
\emph{external} and the \emph{global fields}. A nonuniform field
provides different influences on the agents, while the interaction
with uniform fields is shared by all the elements in the system.
The \emph{local field} (spatially nonuniform) is less efficient
than uniform fields in promoting the formation of multiple
domains, and therefore order is maintained for a larger range of
values of $B$ when interacting with a \emph{local field}.

\section{Effects of an interacting Field for $q>q_c$}

When there are no additional interacting fields ($B=0$), the
system always freezes into disordered states for $q>q_c$. Figure~4
shows the order parameter $g$ as a function of the probability $B$
for the three types of fields. The effect of a field for $q>q_c$
depends on the magnitude of $B$. In the three cases we see that
for $B\rightarrow 0$, $g$ drops to values below the reference line
corresponding to its value when $B=0$. Thus, the limit
$B\rightarrow 0$ does not recover the behavior of the model with
only local nearest-neighbor interactions. The fact that for $B
\rightarrow 0$ the interaction with a field increases the degree
of order in the system is related to the non-stable nature of the
inhomogeneous states in Axelrod's model. When the probability of
interaction $B$ is very small, the action of a field can be seen
as a sufficient perturbation that allows the system to escape from
the inhomogeneous states with frozen dynamics. The role of a field
in this situation is similar to that of noise applied to the
system, in the limit of vanishingly small noise rate \cite{Maxi2}.

The drop in the value of $g$ as $B \rightarrow 0$ from the
reference value ($B=0$) that takes place for the \emph{local
field} in Fig.~4 is more pronounced than the corresponding drops
for uniform fields. This can be understood in terms of a greater
efficiency of a nonuniform field as a perturbation that allows the
system to escape from a frozen inhomogeneous configuration.
Increasing the value of $B$ results, in all three types of fields,
in an enhancement of the degree of disorder in the system, but the
\emph{local field} always keeps the amount of disorder, as
measured by $g$, below the value obtained for $B=0$. Thus a
\emph{local field} has a greater ordering effect than both the
global and the \emph{external field}s for $q>q_c$.

\begin{figure}[t]
\includegraphics[width=0.45\textwidth,angle=0]{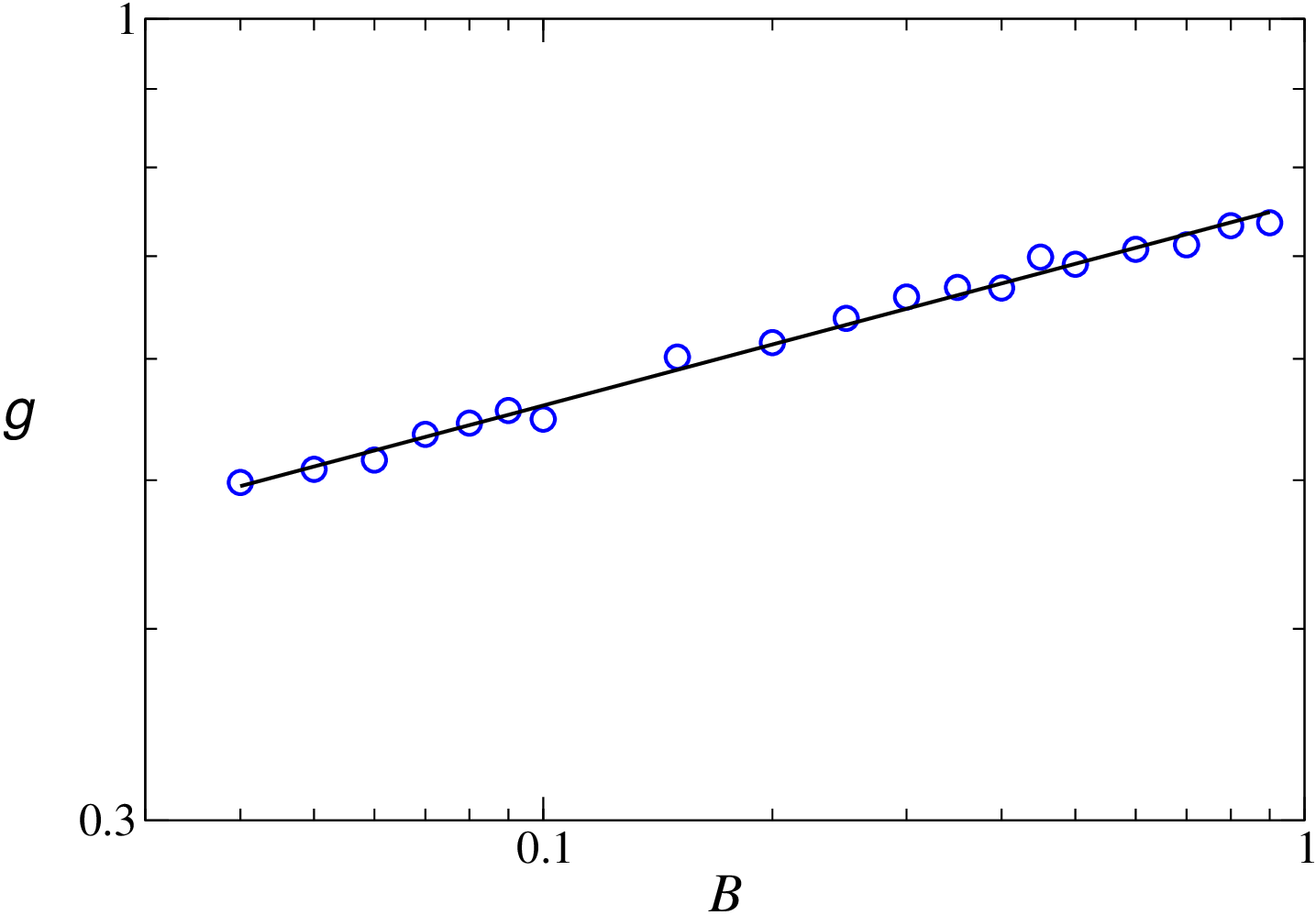}
\caption{Scaling of the order parameter $g$ with the coupling strength to
the \emph{global field} $B$. The slope of the fitting straight line is
$\beta=0.13 \pm 0.01$. Parameter value $q=30>q_c$.}
\end{figure}

The behavior of the order parameter $g$ for larger values of $B$
can be described by the scaling relation $g \sim B^\beta$, where
the exponent $\beta$ depends on the value of $q$. Figure~5 shows a
log-log plot of $g$ as a function of $B$, for the case of a
\emph{global field}, verifying this relation. This result suggests
that $g$ should drop to zero as $B\rightarrow 0$. The partial
drops observed in Fig.~4 seem to be due to finite size effects for
$B\rightarrow 0$. A detailed investigation of such finite size
effects is reported in Fig.~6 for the case of the \emph{global
field}. It is seen that, for very small values of $B$, the values
of $g$ decrease as the system size $N$ increases. However, for
values of $B \gtrsim 10^{-2}$, the variation of the size of the
system does not affect $g$ significantly.

\begin{figure}[t]
\includegraphics[width=0.45\textwidth]{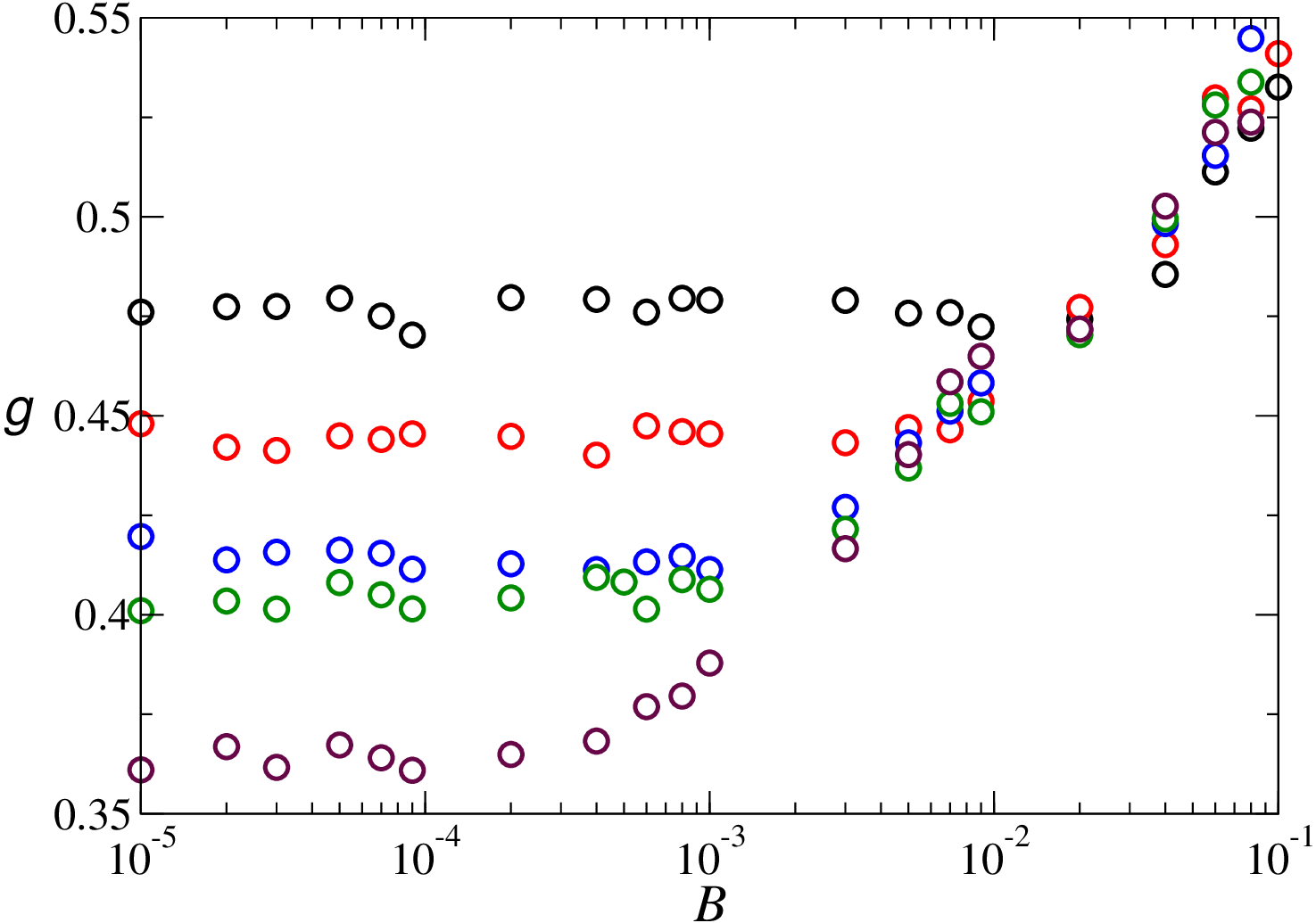}
\caption{Finite size effects at small values of the strength $B$
of a \emph{global field}. Order parameter $g$ as a function of $B$
is shown for system sizes $N=20^2$, $30^2$, $40^2$, $50^2$, $70^2$
(from top to bottom). Parameter value $q=30$.}
\end{figure}

Figure~7 displays the dependence of $g$ on the size of the system
$N$  when $B \rightarrow 0$ for the three interaction fields being
considered. For each size $N$, a value of $g$ associated with each
field was calculated by averaging over the plateau values shown in
Fig. 6 in the interval $B \in [10^{-5},10^{-3}]$. The mean values
of $g$ obtained when $B=0$ are also shown for reference. The order
parameter $g$ decreases for the three fields as the size of the
system increases; in the limit $N \rightarrow \infty$ the values
of $g$ tend to zero and the system becomes homogeneous in the
three cases. For small values of $B$, the system subject to the
\emph{local field} exhibits the greatest sensitivity to an
increase of the system size, while the effect of the constant
\emph{external field} is less dependent on system size. The
ordering effect of the interaction with a field as $B \rightarrow
0$ becomes more evident for a local (nonuniform) field. But, in
any case, the system is driven to full order for $B \rightarrow 0$
in the limit of infinite size by any of the interacting fields
considered here.

\section{Summary and conclusions}

We have analyzed a nonequilibrium lattice model of locally
interacting elements and subject to additional interacting fields.
The state variables are described by vectors whose components take
discrete values. We have considered the cases of a constant
\emph{external field}, a \emph{global field},  and a \emph{local
field}. The interaction dynamics, based on the similarity or
overlap between vector states,  produces several nontrivial
effects in the collective behavior of this system. Namely, we find
two main effects that contradict intuition based on the effect of
interacting fields in equilibrium systems where the dynamics
minimizes a potential function. First, we find that an interacting
field might disorder the system: For parameter values for which
the system orders due to the local interaction among the elements,
there is a threshold value $B_c$ of the probability of interaction
with a field. For $B>B_c$ the system becomes disordered. This
happens because there is a competition between the consequences of
the similarity rule applied to the local interactions among
elements, and applied to the interaction with the field. This
leads to the formation of domains and to a disordered system. A
second effect is that, for parameter values for which the dynamics
based on the local interaction among the elements leads to a
frozen disordered configuration, very weak interacting fields are
able to order the system. However, increasing the strength of
interaction with the field produces growing disorder in the
system. The limit $B \rightarrow 0$ is discontinuous and the
ordering effect for $B<<1$ occurs because the interaction with the
field acts as a perturbation on the non stable disordered
configurations with frozen dynamics appearing for $B=0$. In this
regard, the field behaves similarly to a random fluctuation acting
on the system, which always induces order for small values of the
noise rate \cite{Maxi2}.

\begin{figure}[t]
\includegraphics[width=0.45\textwidth]{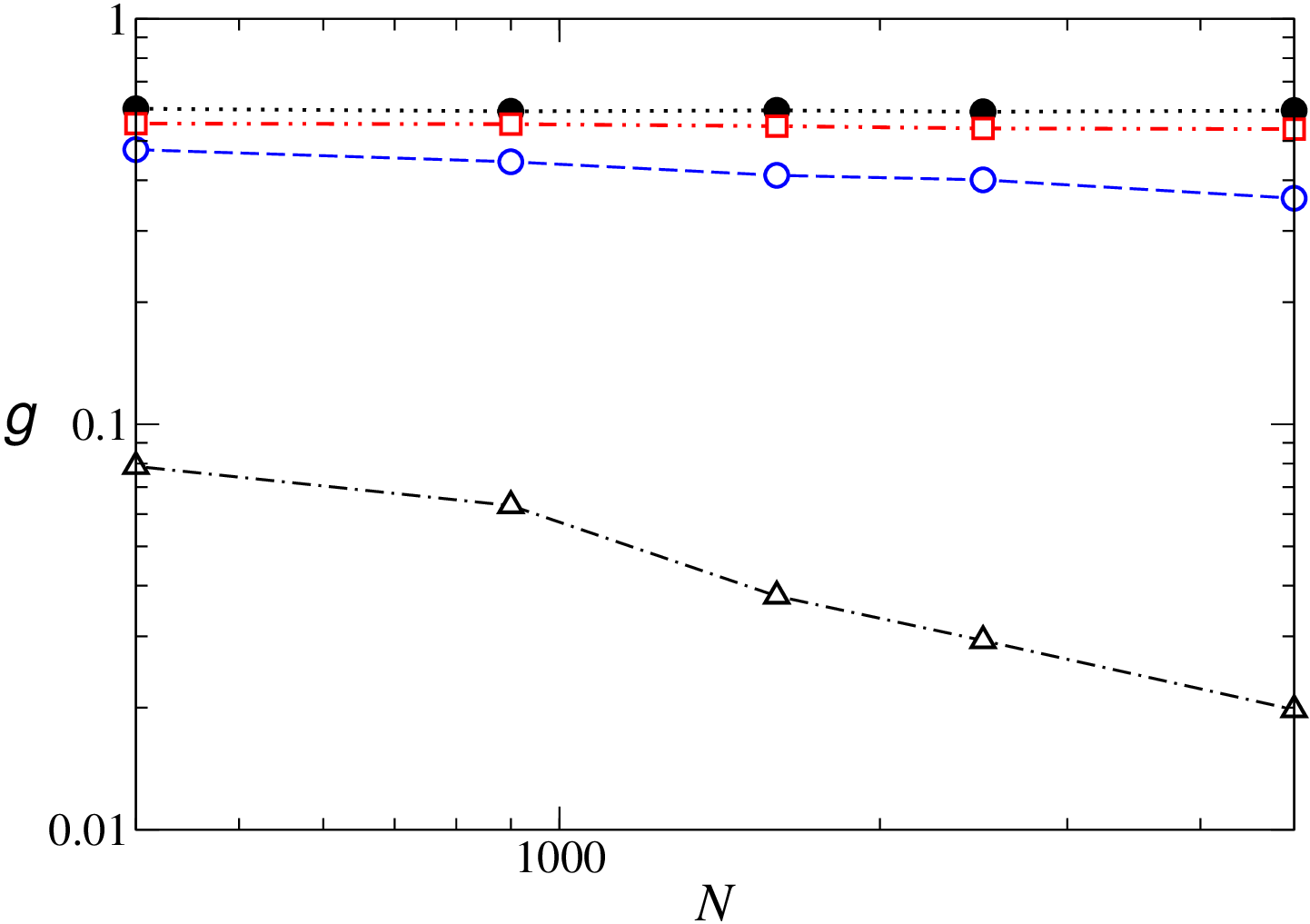}
\caption{ Mean value of the order parameter $g$ as a function of
the system size $N$ without field ($B=0$, solid circles), and with an
\emph{external} (squares), \emph{global} (circles) and
\emph{local} \emph{field} (triangles). Parameter value
$q=30$.  }
\end{figure}

These results are summarized in Fig.\ 8 which shows, for different
values of $B$, the behavior of the order parameter $\langle
S_{max}\rangle/N$ previously considered in Fig. 1. For small
values of $B$, the interaction with a field can enhance order in
the system: for $q<q_c$  interaction with a field preserves
homogeneity, while for $q>q_c$ it causes a drop in the degree of
disorder in the system. In an effective way the nonequilibrium
order-disorder transition is shifted to larger values of $q$ when
$B$ is non-zero but very small. For larger values of $B$ the
transition shifts to smaller values of $q$ and the system is
always disordered in the limiting case $B \rightarrow 1$. This
limiting behavior is useful to understand the differences with
ordinary dynamics leading to thermal equilibrium in which a strong
field would order the system. In our nonequilibrium case, the
similarity rule of the dynamics excludes the interaction of the
field with elements with zero overlap with the field. Since the
local interaction among the elements is negligible in this limit,
there is no mechanism left to change situations of zero overlap
and the system remains disordered. We have calculated, for the
three types of field considered, the corresponding
boundary in the space of parameters $(B,q)$ that separates the
ordered phase from the disordered phase. In the case of a constant
\emph{external field}, the ordered state in this phase diagram
always converges to the state prescribed by the constant field
vector. The nonuniform \emph{local field} has a greater ordering
effect than the uniform (\emph{global} and constant
\emph{external}) fields in the regime $q>q_c$.  The range of
values of $B$ for which the system is ordered for $q<q_c$ is also
larger for the nonuniform \emph{local field}.

\begin{figure}[t]
\includegraphics[width=0.45\textwidth]{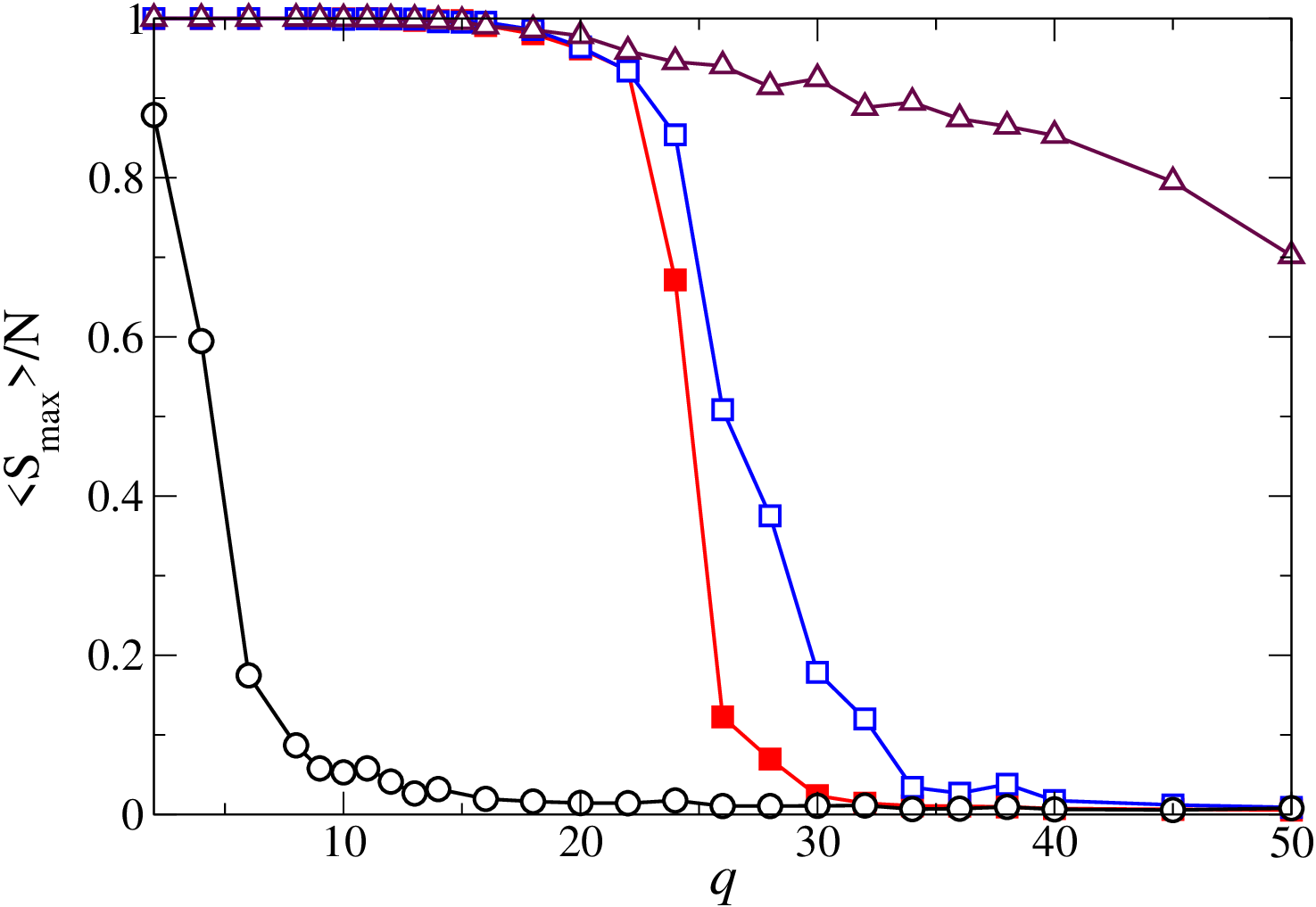}
\caption{ Influence of the interacting field on the nonequilibrium
order-disorder transition as described by the order parameter
$\langle S_{max}\rangle/N$. Results are shown for $B=0$ (solid squares),
a \emph{global} ($B=10^{-5}$ (empty squares),
$B=0.3$ (circles)) and a \emph{local}
($B=10^{-5}$ (triangles))  \emph{field}. Parameter value $F=3$. }
\end{figure}

In spite of the differences mentioned between uniform and
nonuniform fields, it is remarkable that the collective behavior
of the system displays analogous phenomenology for the three types
of fields considered, although they have different nature. At the
local level, they act in the same manner, as a ``fifth" effective
neighbor whose specific source becomes irrelevant. In particular,
both uniform fields, the \emph{global} coupling and the
\emph{external field}, produce very similar behavior of the
system. Recently, it has been found that, under some
circumstances, a network of locally coupled dynamical elements
subject to either global interactions or to a uniform external
drive exhibits the same collective behavior \cite{CPP,PC}. The
results from the present nonequilibrium lattice model suggest that
collective behaviors emerging in autonomous and in driven
spatiotemporal systems can be equivalent in a more general
context.

In the context of Axelrod's model for the dissemination of culture
\cite{Axelrod} the interacting fields that we have considered can
be interpreted as different kinds of mass media influences acting
on a social system. In this context, our results suggest that
both, an externally controlled mass media or mass media that
reflect the predominant cultural trends of the environment, have
similar collective effects on a social system. We found the
surprising result that, when the probability of interacting with
the mass media is sufficiently large, mass media actually
contribute to cultural diversity in a social system, independently
of the nature of the media. Mass media is only efficient in
producing cultural homogeneity in conditions of weak broadcast of
a message, so that local interactions among individuals can be
still effective in constructing some cultural overlap with the
mass media message. Local mass media appear to be more effective
in promoting uniformity in comparison to global, uniform
broadcasts.

Future extensions of this work should include the consideration of
noise and complex networks of interaction.

\section*{Acknowledgments}
J.\ C. \ G-A., V.\ M.\ E. and M.\ SM acknowledge financial support
from MEC (Spain) through projects CONOCE2 (FIS2004-00953) and
FIS2004-05073-C04-03. M.\ G.\
C.\ and J.\ L.\ H.\ acknowledge support from C.D.C.H.T.,
Universidad de Los Andes (Venezuela) under grant No.
C-1285-04-05-A. K.\ K.\ acknowledges support from DFG
Bioinformatics Initiative BIZ-6/1-2 and from Deutscher
Akademischer Austausch Dienst (DAAD).

\end{document}